\documentclass[conference]{IEEEtran}
\usepackage{cite,citesort}
\usepackage{amsmath,amssymb,amsfonts}
\usepackage{algorithmic}
\usepackage{ntheorem}
\usepackage{bbm}
\newtheorem{theorem}{Theorem}
\usepackage[ruled,vlined]{algorithm2e}
\usepackage{graphicx}
\usepackage{textcomp}
\usepackage{xcolor}
\usepackage{float}
\def\BibTeX{{\rm B\kern-.05em{\sc i\kern-.025em b}\kern-.08em
    T\kern-.1667em\lower.7ex\hbox{E}\kern-.125emX}}

\begin{document}

\title{Decentralized Collaborative Video Caching in\\ 5G Small-Cell Base Station Cellular Networks}

\author{\IEEEauthorblockN{Shadab Mahboob, Koushik Kar}
\IEEEauthorblockA{Electrical, Computer and Systems Engineering\\ Rensselaer Polytechnic Institute\\ Troy, NY 12180. \\ Email: \{mahbos, kark\}@rpi.edu\vspace{-10cm}}
\and
\IEEEauthorblockN{Jacob Chakareski}
\IEEEauthorblockA{College of Computing\\ New Jersey Institute of Technology\\ Newark, NJ 07102. \\ Email: jacob.chakareski@njit.edu}\vspace{-1cm}}

\maketitle
\thispagestyle{plain}
\pagestyle{plain}

\begin{abstract}
 We consider the problem of video caching across a set of 5G small-cell base stations (SBS) connected to each other over a high-capacity short-delay back-haul link, and linked to a remote server over a long-delay connection. Even though the problem of minimizing the overall video delivery delay is NP-hard, the Collaborative Caching Algorithm (CCA) that we present can efficiently compute a solution close to the optimal, where the degree of sub-optimality depends on the worst case video-to-cache size ratio. The algorithm is naturally amenable to distributed implementation that requires zero explicit coordination between the SBSs, and runs in $O(N + K \log K)$ time, where $N$ is the number of SBSs (caches) and $K$ the maximum number of videos.  We extend CCA to an online setting where the video popularities are not known a priori but are estimated over time through a limited amount of periodic information sharing between SBSs. We demonstrate that our algorithm closely approaches the optimal integral caching solution as the cache size increases. Moreover, via simulations carried out on real video access traces, we show that our algorithm effectively uses the SBS caches to reduce the video delivery delay and conserve the remote server's bandwidth, and that it outperforms two other reference caching methods adapted to our system setting. 
\end{abstract}


\section{Introduction}
Recently, the Internet has witnessed deployment of a variety of video streaming applications, as well as tremendous growth in video traffic. Applications such as YouTube, Netflix, Amazon Prime Video, Hulu, and Sling TV are contributing to a large part of our daily Internet bandwidth consumption. Live video streaming and video-on-demand (VoD) services are growing, sharing of video news and messages through social networking applications like Facebook and WhatsApp is increasing steadily, and news readers are increasingly utilizing video feeds for their daily news. A recent Cisco study \cite{Cisco2019} finds that Internet video traffic has been growing annually at 33\% and will constitute about 82\% of all IP traffic by 2022.

The growth in video traffic is also accompanied by significant changes in video access patterns in recent years. 
Firstly, the quality and bit-rate of videos are steadily increasing, with growing availability of Ultra High Definition (UHD) or 4K video streams that occupy more than double the HD video bit rate.  
Secondly, the difference between `live’ and `stored’ video is blurring. As more users expect VoD capability for TV shows -- where Internet-based delivery mechanisms are making steady inroads -- access to TV shows gets staggered over time. Nevertheless, the access patterns of a show are often highly correlated temporally, i.e., close to each other in time, based on when it is posted online. 
Finally, there is considerable growth in video viewing over wireless, over both WiFi and cellular technologies.  Emerging super-fast access technologies such as 802.11ac/802.11ax and 5G are accelerating this growth. There is also increased video viewing on mobile devices such as smartphones and tablets, which are expected to contribute to about 50\% of the Internet traffic in a few years \cite{Cisco2019}. 

These trends indicate an increased importance of caching videos close to the users, to reduce the access delay and network/server congestion. Increasing deployment of 5G access points is expected over the next decade, making these access points (or \textit{Small-cell Base Stations (SBS)}) natural candidates for hosting such caches. The range (coverage area) of these SBSs, and the cache sizes that can be included in them, are expected to be small. 
At the same time, in many of the deployment scenarios (malls, office buildings, campuses etc.), such SBSs may be deployed in large numbers, and be connected with each other other over a fast local area network such as high-speed Ethernet. This motivates pooling resources of multiple such SBSs, and using them to collaboratively cache videos for access by users covered by a cache pool.

In this paper, we consider the problem of video caching in a wireless edge network comprising of multiple small-cell base stations (SBSs) that are connected to each other over a high-capacity low-delay local area network. The SBSs host small video caches but can exchange videos with each other over the local network; or they can fetch videos from a remote server over a long-delay Internet path. In this setup, the problem of minimizing the overall video playout delay is NP-hard due to packing-type integrality constraints; even if the integrality constraints are relaxed, the problem is a concave minimization problem which could be NP-hard in general. Despite these facts, we utilize the specific structure of our problem to develop an efficient algorithm that computes a close-to-optimal solution, where the degree of sub-optimality depends on the worst case video-to-cache size ratio. More specifically, our algorithm is naturally amenable to distributed implementation and runs in $O(N + K \log K)$ time, where $N$ is the number of SBSs (caches) and $K$ the maximum number of videos. 
We also extend this algorithm to a dynamic setting where the video popularities are not known a priori but are estimated over time. The distributed, online implementation \textit{does not require any explicit coordination between the SBSs}, as long as an ordering (tie-breaking) rules between the caches and the videos are pre-determined, and information on the video requests from users are periodically shared between the SBSs.
We show via numerical experiments on a small number of caches that our algorithm approaches the optimal (integral) caching solution when the cache size is large relative to the individual video sizes. Simulations conducted on real video access traces demonstrate the performance trade-offs between video playout delay and local and remote bandwidth used, and the effect of popularity estimation parameters and re-optimization intervals on the overall caching performance of the system. We show that when there is significant temporal correlation in video access patterns across the caches, our algorithm is able to effectively use collaboration between the SBS cache pool to reduce the video playout delay and save the remote server's bandwidth. 

\section{Related Work}
\label{sec:related-work}

The general caching problem has attracted considerable attention in the last decade due to the emergence of content-centric networks, e.g., \cite{6655114,7636877}. 
These studies mainly focused on either lowering the bandwidth consumption during the peak traffic times or the content delivery delay by developing efficient algorithms. 
With the advent of cellular edge networks, SBSs have become the most suitable candidates for caching with reduced latency, cost and energy consumption \cite{7445129}. 
In \cite{6883600}, content delivery delay is minimized for a single SBS by formulating this is as a knapsack problem \cite{10.5555/98124} based on derived content popularity. Authors in \cite{6600983} analyze ``femtocaching'', where small wireless caching helpers with limited size and coverage area reduce the content delivery delay. 
Hierarchical coded caching in networks with multiple layers of caches is introduced in \cite{7458151}. As the SBSs are deployed far from each other in traditional wireless cellular networks, these schemes mainly focus on caching optimization on a scale of a single SBS without considering possible collaboration among SBSs. Densely located SBSs in 5G networks make collaborative caching among SBSs feasible.

To this end, a number of cooperative caching strategies for cellular networks have been proposed using primarily optimization approaches, considering different objective functions such as overall delay, cost and revenue, together with cache capacity constraints. The study in \cite{7179394} maximizes a total reward objective for an ISP in a collaborative manner, given limited caching space at each SBS, and formulates strong approximation algorithms for both, coded and uncoded data cases. In a follow-up study, joint caching, routing, and channel selection is investigated using large-scale column generation optimization with tight approximation guarantees \cite{KhreishahCG:15}. In \cite{7815021}, the aggregated storage and download cost for caching is minimized for both limited and unlimited caching spaces by devising a near optimal greedy algorithm. 
However, 
implementing complex centralized algorithms may be infeasible in practice, and 
a simple distributed, adaptive algorithm that requires minimal coordination between the SBSs is what is practically desirable.

\section{Model and Formulation}
\label{sec:model}
\begin{figure}[htb]
   \vspace{-0.1in}
    \centering
    \includegraphics[width=.8\linewidth]{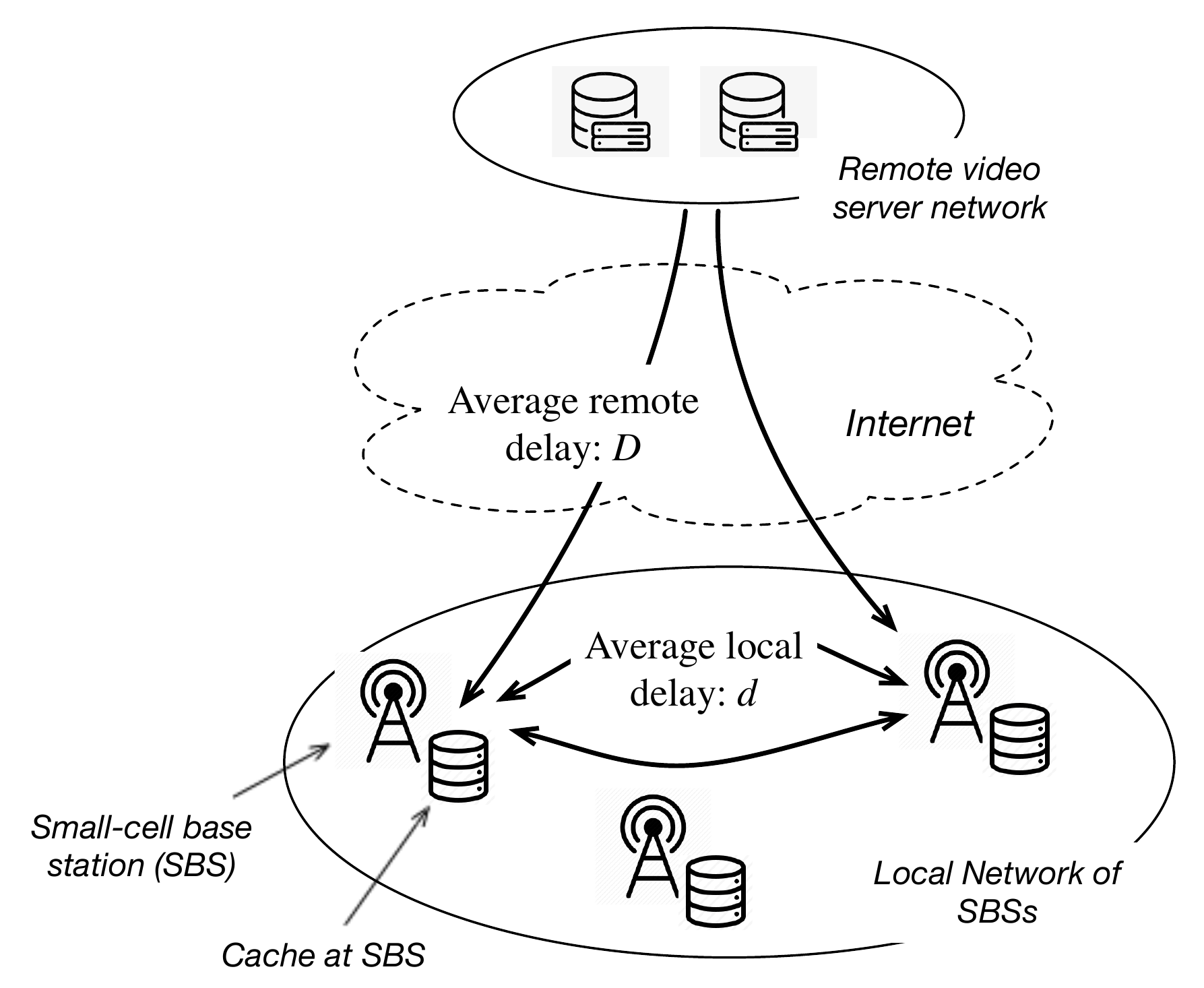}
    \caption{Illustration of our System Model.}
    \vspace{-0.05in}
    \label{fig:model}
\end{figure}

We consider an edge network comprised of a set of $N$ SBSs, indexed as $i=1, \cdots, N$. There is a set of $K$ videos, indexed as $k=1, \cdots, K$, which may be downloaded from one or more remote servers. 
SBS $i$ is associated with cache space $C_i$, which it uses to selectively cache some of the videos. The SBSs are connected to a high-speed local network over which they can exchange videos with each other. Each end-user is assumed to be associated with one of the SBSs at any given time, although that association may change over time due to user mobility. If the requested video is available at the SBS (cache) that the user is associated with, the video is served to the user with minimal playout delay.\footnote{For ease of exposition, we take this delay to be zero. There is no loss of generality here, as assuming that this delay on an average is a positive number $\delta$ does not affect our algorithm or its analysis.} Otherwise, the video is either obtained from one of the other SBSs in the local network, or is downloaded from the remote server(s) if videos are not available in the local caches. If the video is present in one of the other local SBS caches, we assume an average playout delay of $d$; otherwise (i.e., if the video is to be fetched from the remote server(s)), the average playout delay is average playout delay is $D > d$.\footnote{Note that we consider the video \textit{playout} delay and not the video delivery delay, and therefore this delay is assumed independent of the video size. The video playout delay, one of the most important factors affecting Quality of Experience (QoE), is primarily a function of the quality of the connection (path) (such as bandwidth, round trip time) over which the video is delivered.} In the rest of this paper, the term `delay' refers to video playout delay, unless mentioned otherwise. The system model is illustrated in Figure~\ref{fig:model}. 


Let the popularity and size of video $k$ be represented by $\pi_k$ and $s_k$, respectively. Note that the video popularity is considered independent of the SBS (cache) $k$. Since users will typically be mobile across the coverage area of the SBSs, we reasonably assume that the same popularity vector (which represents the access rates of the videos across the entire population of served users) would apply to all SBSs. Given $\mathbf{\pi} = (\pi_k, k=1, \cdots, K)$, and letting $x_{i,k}$ be a binary variable indicating if video $k$ is cached at SBS $i$, our goal is to minimize the overall video playout delay, expressed as 
$$\sum_k \pi_k  \bigg[ \sum_i \Big( \left(\max_{i'} x_{{i'},k}-x_{i,k}\right)  d \ + \left(1- \max_{i'} x_{{i'},k}\right) D \Big) \bigg].$$
Note that  $\left(\max_{i'} x_{{i'},k}-x_{i,k}\right)$ equals 1 only when video $k$ is cached locally (in one of the SBS caches) but not at cache $i$, and zero otherwise. Therefore, the term $\left(\max_{i'} x_{{i'},k}-x_{i,k}\right) d$ represents the additional video delay incurred if requested video $k$ is not cached at SBS $i$ but has to be fetched from one of the other SBSs. in On the other hand, the term $\left(1- \max_{i'} x_{{i'},k}\right)$ is 1 only if video $k$ is not cached locally at all (and therefore must be fetched from a remote server incurring an additional average delay cost of $D$), and zero otherwise. Since the above objective is equivalent to maximizing $\sum_k \pi_k  \bigg[ d \sum_i x_{i,k}  \ + \  (D-d)  \sum_i \max_{i'} x_{{i'},k} \bigg]$, we define the \textit{Collaborative Caching Problem (CCP)} as
\vspace{-0.05in}
\begin{eqnarray}
\mbox{maximize  } \sum_k \pi_k  \bigg[ d \sum_i x_{i,k}  \ + \  (D-d)  \sum_i \max_{i'} x_{{i'},k} \bigg],
\label{eq:obj}
\end{eqnarray}
\vspace{-0.1in}
\begin{eqnarray}
\mbox{subject to} & \sum_k x_{i,k} s_k  \leq \  C_i, & \forall i, \label{eq:constant-cap}\\
& x_{i,k} \in \  \{0,1\}, & \forall i, \forall k. \label{eq:constant-int}
\end{eqnarray}




Next we mention a few points about the structure and complexity of the CCP formulation described in (\ref{eq:obj})-(\ref{eq:constant-int}), towards motivating the solution approach that we will present in the next section. The complexity of the problem comes from two aspects: (a) the non-linearity (non-convexity) of the objective function (\ref{eq:obj}); and (b) the integrality (binary nature) of the optimization variables in (\ref{eq:constant-int}). The constraints (\ref{eq:constant-cap})-(\ref{eq:constant-int}) represents integral packing type constraints which makes the problem NP-hard. For a single SBS, the problem reduces to the 0-1 knapsack problem \cite{10.5555/98124}. While the binary knapsack problem is NP-hard, is can be solved in pseudo polynomial time using dynamic programming; several heuristic approaches are also known to work well in practice. However, most of these approaches are not easily amenable to distributed implementation with very low coordination and message exchange complexity, which is highly desirable in our setting. 

Towards addressing (a), we note that since $D > d$, when the integrality constraints (\ref{eq:constant-int}) are relaxed, the CCP posed in (\ref{eq:obj})-(\ref{eq:constant-int}) represents a \textit{concave minimization} problem over a polyhedral set \cite{Benson1995}. While such problems are NP-hard in general, the $\max$ terms in the objective function can be replaced by a set of linear terms and additional linear constraints. However, this not only results in additional variables and constraints, it does not help towards developing distributed solutions with low coordination message complexity. 




In our algorithm that we describe next (in Section~\ref{sec:algo}), both (a) and (b) are addressed, and the resulting solution closely approximates the optimal integral solution of the problem, when the SBS cache capacities are sufficiently large compared to the individual video sizes (or units in which the videos are cached)\footnote{If video sizes are large (like HD/UHD quality movies), caches may store the beginning few minutes of each video (instead of the entire video, which may be large), seeking to reduce the initial playout delay, while the rest of the video is streamed directly from the server to the user. In other cases where the video size is large, such as in 360-degree videos, the video can be cached in units of small-size tiles \cite{Chakareski:18}. Therefore, our assumption on the ratio of the video caching unit to the cache size being small generally holds true.}.Further, the algorithm can be implemented is a distributed manner with \textit{no explicit coordination} between the SBSs, as long certain tie-breaking rules are agreed upon in advance, and information about video requests from users are periodically shared between SBSs to enable estimation of the popularity vector $\pi$.

\section{Algorithm and Analysis}
\label{sec:algo}

Towards developing intuition behind the algorithm and its optimality, we first present the \textit{Collaborative Caching Algorithm (CCA)} for the special case of unit size videos, with the cache sizes (possibly different from each other) being integral multiples of this unit video size. This special case admits a simpler algorithm that can be described and illustrated easily, as well as a simpler optimality proof that still captures the essence of the argument. We then extend this algorithm and analysis to the general case of arbitrary video sizes (in addition to cache sizes being arbitrary), for which we argue that our algorithm is optimal in an approximate (asymptotic) sense.

\subsection{Collaborative Caching for Unit Video Sizes}
In this special case, $s_k=1 \ \forall k$; also, $C_i = m_i$ for some positive integer $m_i \ \forall i$. This implies that there is no loss due to fragmentation of the videos. The formulation of CCP remains the same as that in (\ref{eq:obj})-(\ref{eq:constant-int}) except that (\ref{eq:constant-cap}) is reduces to $\sum_{k \in \mathcal{K}} x_{i,k}  \leq \  C_i \ \forall i$. 

\begin{figure*}[t]
    \centering
    \includegraphics[width=\linewidth]{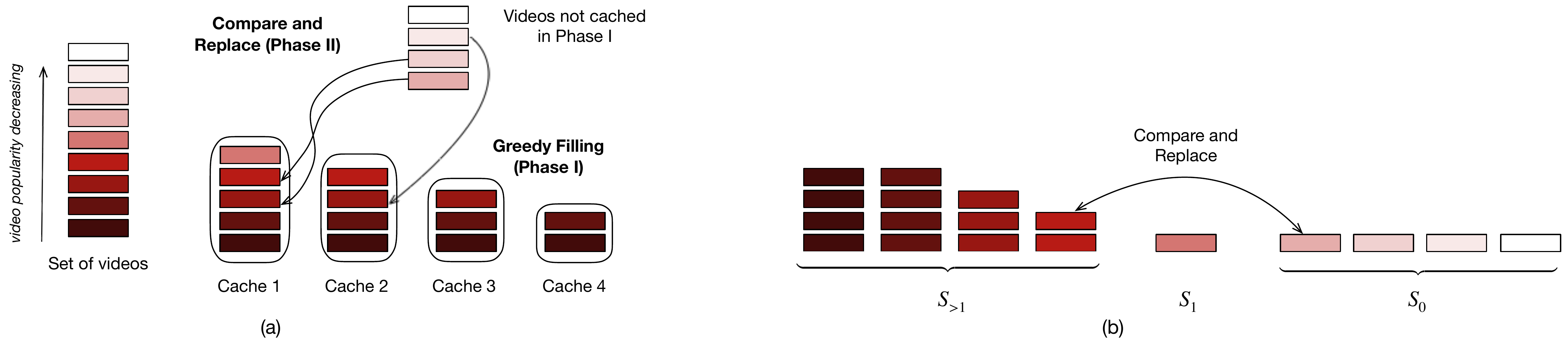}
    \vspace{-0.3in}
    \caption{Illustration of CCA for unit video sizes.}
    \label{fig:algo}
    \vspace{-0.15in}
\end{figure*}

Without loss of generality, assume that the videos are indexed in decreasing order of their popularity values (ties broken arbitrarily), i.e., $\pi_k \geq \pi_{k'}$ when $k<k'$. Let the number of cached copies of video $k$ during the run of the algorithm be denoted by $n_k$. The algorithm works in two phases:

\noindent \textbf{Phase I (Greedy Filling):} In this phase, each cache is filled up greedily and independently with videos in the order $1, 2, 3, \cdots$ (decreasing order of popularity) up to the cache size limit. For the case of unit video sizes, this simply means that cache $i$ caches up to video with index $C_i = m_i$. Note that at the end of this phase, the highly popular videos may be cached in multiple (possibly all) caches, whereas videos with low popularity may not be cached at all. In general, a video $k$ is cached in any cache $i$ such that $k \leq m_i$.
Let $S_{>1}, S_1$, and $S_0$ respectively denote the sets of videos that have been cached at multiple caches, at a single cache, or none at all. 
The greedy filling step is illustrated in Figure~\ref{fig:algo}(a). 

\noindent \textbf{Phase II (Compare and Replace):} 
In the second phase, the $n_k$ values (and accordingly, the $S_{>1}, S_1$, and $S_0$ sets) are altered as some of the videos from the set $S_0$ are cached in one of the caches (and therefore move to set $S_1$) replacing videos that are present in more than one cache. This replacement happens if two videos $k_1 \in S_{>1}$ and $k_2 \in S_0$ satisfy a ``popularity ratio test'': $\pi_{k_2}/\pi_{k_1} > d/(ND - (N-1)d)$.\footnote{It will be observed later in our analysis that replacement according to this popularity ratio test improves the objective function in (\ref{eq:obj}).} Thus, in this phase, the $n_k$ values of the videos in $S_{>1}$ may reduce, and the set $S_{>1}$ may shrink as well, as some of the videos in $S_{>1}$ may just have one copy left and therefore move to the set $S_1$. Further, no video in $S_0$ or $S_1$ is cached multiple times in this phase, so no video is added to $S_{>1}$. The set $S_0$ may shrink as well, as some of its videos may be cached (in a single cache at most) and therefore move to $S_1$. The set $S_1$ can only expand, due to videos from both $S_{>1}$ and $S_0$ possibly moving to this set. See the illustration in Figure~\ref{fig:algo}(b). The popularity ratio test is conducted by picking $k_1$ to be the video with the largest index in $S_{>1}$ and $k_2$ as the video with the smallest index in $S_0$; therefore the replacement step can be viewed as the boundary between $S_1$ and $S_0$ moving right by one step; and the boundary between $S_{>1}$ and $S_1$ staying the same, or moving left by one step. The replacement stops (and the algorithm ends) when the $k_1, k_2$ thus picked fails the ratio test, or either $S_{>1}$ or $S_0$ becomes empty (i.e., $k_1$ becomes $0$ or $k_2$ becomes $N+1$).

The algorithm is described in Algorithm~\ref{algo:unit-size_caching}. Assuming the videos are already pre-sorted according to their $\pi_k$ values, Phase II involves up to $K$ compare and replace steps, and therefore has $O(K)$ complexity. The complexity of Phase II is $O(K)$ per cache, or $O(NK)$.
Including the complexity of the sorting step (which is needed for Phase I as well), the complexity of CCA for unit size videos is $O(NK + K \log K)$.

\begin{algorithm}
 \textbf{\underline{Phase I (Greedy Filling)}:}\\
\For{each SBS (cache) $i$}{
Fill up cache $i$ with videos $1, 2, \cdots, m_i$.
 }
 \For{each video $k$}{
$n_k = \{i | k \leq m_i \}$ // number of cached copies of $k$.
 }
 Set $S_{>1} = \{k | n_k>1\}, S_{1} = \{k | n_k=1\}, S_{0} = \{k | n_k=0\}$.\\
 \textbf{\underline{Phase II (Compare and Replace)}:}\\
 Let $k_1$ = last index in $S_{>1}$; $k_2$ = first index in $S_{0}$. \\
 \While{($S_{>1}$ and $S_0$ are both non-empty) \&\& ($\frac{\pi_{k_2}}{\pi_{k_1}} > \frac{d}{ND - (N-1)d}$)}{
 Replace any one copy of video $k_1$ with $k_2$, and set $n_{k_1} = n_{k_1} - 1$ and $n_{k_2} = 1$.\\
  Set $S_0 = S_0 - {\{k_2\}}$, $S_1 = S_1 + {\{k_2\}}$, and $k_2 = k_2+1$.\\
  \If{($n_{k_1} == 1$)} {Set $S_{>1} = S_{>1} - \{k_1\}; S_{1} = S_{1} + \{k_1\}$, and $k_1 = k_1 - 1$.}
 }
\caption{Collaborative Caching Algorithm (CCA): \textit{Unit Video Sizes} (\textbf{Input:} $d,D,\pi$)}
\label{algo:unit-size_caching}
\end{algorithm}

Our optimality claim for the special case of unit video sizes is stated in Theorem~\ref{th:unit-size} (proof in Appendix).

\begin{theorem}
\label{th:unit-size}
If $s_k = 1$ for all videos $k$, and all cache sizes $C_i$ are integral, then Algorithm~\ref{algo:unit-size_caching} computes an optimal solution of CCP as posed in  
(\ref{eq:obj})-(\ref{eq:constant-int}).
\end{theorem}

\subsection{Collaborative Caching for Arbitrary Video Sizes}
For arbitrary video sizes, the Collaborative Caching Algorithm (CCA) works along similar lines as Algorithm~\ref{algo:unit-size_caching}, but with the following differences. Firstly, in both Phase I (Greedy Filling) and Phase II (Compare and Replace), we ignore the integrality constraints, and allow videos to be filled or replaced \textit{fractionally} if needed, so that no cache space is wasted. Secondly, popularity values $\pi_k$ are replaced by \textit{popularity density} values $w_k = \pi_k/s_k$, and the video indices are sorted in the decreasing order of the $w_k$ values, i.e., $w_k \geq w_{k'}$ when $k < k'$. Finally, there is a final Rounding phase (Phase III) of the algorithm where the space allocated to videos in $S_1$ are reallocated to the same videos, but in a way that satisfies the integrality constraints. At the same time, any fractional videos included in $S_{>1}$ are dropped. Therefore, the rounding step ensures that the final solution contains only full videos. Further details on the three phases of the algorithm is provided next, and
the Collaborative Caching Algorithm (CCA) for this general case is fully described in Algorithm~\ref{algo:arbitrary-size_caching}. 

\noindent \textbf{Phase I (Greedy Filling (\textit{fractional})):} Note that in this phase, cache $i$ stores up to video $m_i$, where $m_i$ satisfies $\sum_{k=1}^{m_i-1} s_k < C_i$ and $\sum_{k=1}^{m_i} s_k \geq C_i$. Thus videos $1, \cdots, m_i-1$ are fully included, but only a part of video $m_i$ may be included to fill up the cache space.


\noindent \textbf{Phase II (Compare and Replace (\textit{fractional})):} 
In this phase, we replace $n_k$ by $y_k$, the aggregate size (including fractional) of video $k$ as cached across all SBSs. Thus, if video $k$ is cached fully in only one cache, $y_k = s_k$. The sets $S_{>1}, S_1$, and $S_0$ are defined accordingly, as shown in Algorithm~\ref{algo:arbitrary-size_caching}. In Phase II, ``extra'' copies of video $k_1 \in S_{>1}$ are replaced by video $k_2 \in S_0$ (even though this replacement may only be fractional), if the popularity density values  of videos $k_1, k_2$ satisfy a ratio test. 


\noindent \textbf{Phase III (Rounding):} The rounding phase collects all the space allocated to videos in $S_1$ across the different caches, and reallocates those videos in that space sequentially. Note that the Phase II may have split the single copy of a video $k$ (belonging to $S_1$) across multiple caches, and the rounding step collects all that and uses it to place video $k$ in a single cache. At any point the next cache among those that have available space (given by the set $R$) is chosen. If this cache (say $i$) does not have enough remaining space to accommodate the video (say $\kappa$), i.e., $c_i < \kappa$, then video $k$ is dropped. This ensures only full size videos in $S_1$ remain in the final solution. Any fractional videos included in $S_{>1}$ and $S_{0}$ are also dropped. Therefore, the rounding phase ensures that the integrality constraints (\ref{eq:constant-int}) are satisfied by the final solution.

The complexity of Phases I and II in Algorithm~\ref{algo:arbitrary-size_caching} is similar to those of Algorithm~\ref{algo:unit-size_caching}. The Rounding phase takes an additional $O(N+K)$ time. Therefore, the overall time complexity of CCA remains the same as that in the special case, i.e., $O(NK + K \log K)$.

\begin{algorithm}
Define $w_k = \pi_k/s_k \ \forall k$, and reorder the video indices $k$ in the decreasing order of the $w_k$ values.\\
 \textbf{\underline{Phase I (Greedy Filling (\textit{fractional}))}:}\\
\For{each SBS (cache) $i$}{
Fill up cache $i$ with videos $1, 2, \cdots, m_i-1$ (full), and $m_i$ (possibly fractional).
 }
 \For{each video $k$}{
$y_k$ = aggregate size of all cached copies of $k$ (including fractional).
 }
 Set $S_{>1} = \{k | y_k>s_k\}, S_{1} = \{k | y_k=s_k\}, S_{0} = \{k | y_k<s_k\}$.\\
 \textbf{\underline{Phase II (Compare and Replace (\textit{fractional}))}:}\\
 Let $k_1$ = last index in $S_{>1}$; $k_2$ = first index in $S_{0}$. \\
 \While{($S_{>1}$ and $S_0$ are both non-empty) \&\& ($\frac{w_{k_2}}{w_{k_1}} > \frac{d}{ND - (N-1)d}$)}{
 Replace an amount $r$ (from any cache(s)) of video $k_1$ with $k_2$, where $r = \min\{y_{k_1} - s_{k_1}, s_{k_2} - y_{k_2}\}$. 
Set $y_{k_1} = y_{k_1} - r$ and $y_{k_2} = y_{k_2} + r$.\\
\If{($y_{k_1} == s_{k_1}$)} {Set $S_{>1} = S_{>1} - \{k_1\}; S_{1} = S_{1} + \{k_1\}$, and $k_1 = k_1 - 1$.}
\If{($y_{k_2} == s_{k_2}$)} {Set $S_{0} = S_{0} - \{k_2\}; S_{1} = S_{1} + \{k_2\}$, and $k_2 = k_2 + 1$.}
  }
 \textbf{\underline{Phase III (Rounding)}:}\\
 Remove any fractional videos from $S_{>1}$ and $S_0$. \\
Let $c_i$ = the amount of space allocated to videos in $S_1$ in cache $i$ after Phase II. Set $R = \{i | c_i > 0\}$.\\
Let $\kappa_1$ ($\kappa_2$) be the first (last) video in $S_1$.\\
 \While{($\kappa_1 \leq \kappa_2$)}{
 Pick the first cache $i$ in $R$.\\
 \eIf{($c_i > s_{\kappa_1}$)}{
Cache video $\kappa_1$ in $i$, and set $c_i = c_i - s_{\kappa_1}$.}
{$R = R - \{i\}$.}
 $\kappa_1 = \kappa_1+1$.
 }
\caption{Collaborative Caching Algorithm (CCA): \textit{Arbitrary Video Sizes} (\textbf{Input:} $d,D,\pi, s$)}
\label{algo:arbitrary-size_caching}
\end{algorithm}
\vspace{3mm}

Our optimality claim for the general case (arbitrary video and cache sizes) is stated in Theorem~\ref{th:arbitrary-size} (proof in Appendix).

\begin{theorem}
\label{th:arbitrary-size}
Let $\epsilon = \frac{\max_k s_k}{\min_i C_i}$. Then Algorithm~\ref{algo:arbitrary-size_caching} computes a feasible solution to CCP as posed in (\ref{eq:obj})-(\ref{eq:constant-int}), with an objective function value no less than $(1 - O(\epsilon))$ of the optimum.
\end{theorem}

From Theorem~\ref{th:arbitrary-size}, we note that the approximation factor $(1 - O(\epsilon))$, which approaches 1 when the cache sizes are sufficiently large compared to the individual video sizes (or the units in which the videos are cached), which is a reasonable assumption in practice. 



\subsection{Distributed and Online Implementation}
Algorithm~\ref{algo:arbitrary-size_caching} can be easily implemented in a distributed manner provided the SBSs (caches) agree upon (i) The video popularity vector $\pi$, and how ties are broken; (ii) An ordering between the caches. The video popularity vector will typically be estimated based on the user video requests at different caches. If the request data is periodically shared between caches, this estimation can be done independently at each cache. 
Any ties in the $\pi_k$ values can be broken according to some predetermined rule, such as video id. 
The caches can be indexed simply in the decreasing order of their cache sizes, with ties broken according to any pre-determined manner.

Note that Phase I (Greedy Filling) is a fully decentralized process (requiring no coordination) which caches can carry out fully independently of each other. Phases II and III may appear centralized as they consider the set of all caches at the same time; however, under the conditions (i) and (ii) specified above, each cache can carry out the Phase II computations without any coordination with each other. Each cache will calculate the same overall solution from Phases II and III, but will only cache the videos that it itself is supposed to cache. In other words, to implement Algorithm~\ref{algo:arbitrary-size_caching} no explicit coordination or messaging is required between caches. Further, after the completion of the algorithm, each cache will also automatically know which videos are being cached in the other caches. This helps in requesting videos from those caches, as needed in a practical (online) implementation of the algorithm, as we describe next. Note that since Phase I can be run in parallel across the different caches, the time complexity of the distributed implementation of CCA is $O(N + K \log K)$. 

In a real-life scenario, the video population would not be fixed, and the relative popularities of the videos will also evolve over time. To adapt our algorithm to such scenarios, we can re-estimate the video popularity values periodically, and re-optimize the caching solution accordingly. In our performance evaluation as described in Section~\ref{sec:ExperimentalEvaluation}, we re-estimate and popularity vector $\pi$ periodically, by counting the number of videos requested by users in the local network (across all SBSs) over a time window $W$ ending at the current time. This requires the SBSs (caches) \textit{to share with each other the list of videos requested, but no other information exchange or coordination is required}. Each cache then applies Exponential Weighted Moving Average (EWMA) to calculate the the popularity values to be used in the caching re-optimization. More specifically, the popularity value of video $k$ at the beginning of the $t^{\rm th}$ window, $\pi^t_k$, is calculated as

\begin{equation}
    \pi^{t}_k = (1-\alpha) \pi^{t-1}_k + \alpha\frac{n^{t-1}_k}{W},
\end{equation}
\noindent where $n^t_k$ represents the number of video requests during the $(t-1)^{\rm th}$ window (across all SBSs), and $\alpha$, where $0 \leq \alpha \leq 1$ is an appropriately chosen weighting parameter.

Note that CCA needs to be re-run (and a new optimal solution needs to be re-computed) at the end of each window, assuming that the popularity vector $\pi$ has changed significantly from last time. CCA then reoptimizes the caching solution, which happens in the background. In this reoptimization process, CCA tries to minimize the use of remote bandwidth: thus, if a cache needs to obtain a new video due to this reoptimization, it preferentially gets it from another cache, and contacts the server only if the video is not stored in the local network. We observe (see Section~\ref{sec:ExperimentalEvaluation}) that the amount of bandwidth consumed (both locally and remotely) is only a very small fraction of the total bandwidth consumed, implying that this reoptimization process has very limited overhead. 





\section{Performance Evaluation}
\label{sec:ExperimentalEvaluation}
We evaluated the performance of CCA through simulations on real video request data traces from Netflix and YouTube. The nature of the results was largely similar between the two traces, but the effectiveness of caching was more pronounced with the Netflix dataset, due to its higher degree of temporal correlation across video requests, as intuitively expected. Due to limited space, we only present results for the Netflix dataset. 


\subsection{Dataset}




We use the publicly available Netflix Prize dataset \cite{Bennett07thenetflix} for simulating and analyzing the performance of our algorithm. 
The trace includes the pattern of about 15.5 million video requests from a database of 17770 videos. From this we extract the request pattern for the most popular 3000 videos and use it in our simulations. Since video size information is not provided in the dataset, we assumed a uniform distribution over a range between 2.5 GB and 5 GB for estimating the size of the videos. This range of sizes correspond to about 90 mins to 180 mins of video, assuming about 27.77 MB per min of video. While we have experimented with different values (ratios) of playout delays $d$ and $D$, the results presented in this paper are mostly for  $d = 500$ ms, remote delay $D = 5$ s. These numbers correspond to about $13.5$ secs of initial playout buffering of the video, assuming local network bandwidth of 100 Mbps and remote server bandwidth (i.e., the link bandwidth between the local network and the remote server) of 10 Mbps. 

\subsection{Comparison with the Optimal Solution}
\begin{figure}[htb]
    \centering
    \vspace{-0.2cm}
     \includegraphics[width=0.65\linewidth]{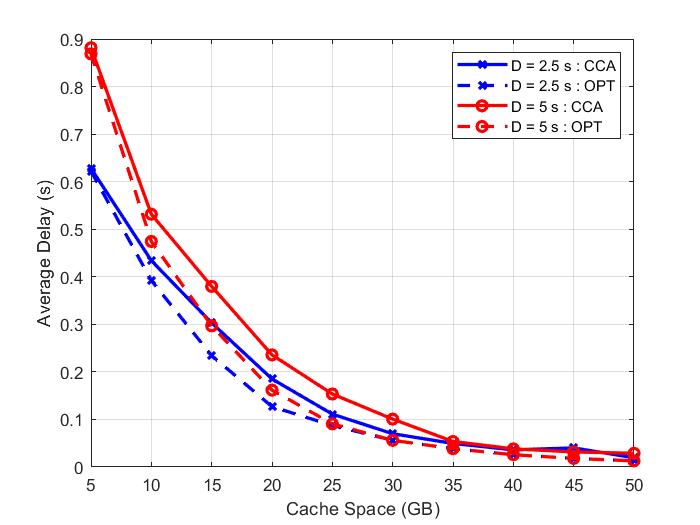}
     \vspace{-0.1cm}
    \caption{Average Delay vs Cache Size at each SBS ($d = 500$ ms).}
    \vspace{-0.1cm}
    \label{fig:cca-vs-opt-cache_size}
\end{figure}

\begin{figure}[htb]
    \centering
    \vspace{-0.2cm}
    \includegraphics[width=0.65\linewidth]{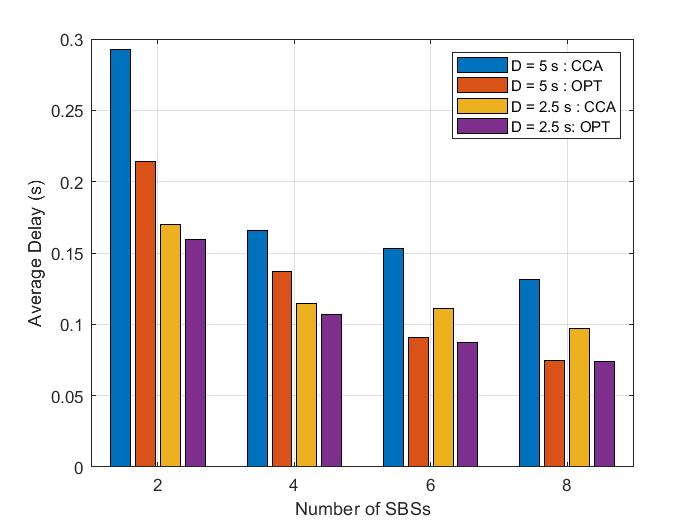}
    \vspace{-0.2cm}
    \caption{Average Delay vs Number of SBSs ($d = 500$ ms in all cases).}
    \vspace{-0.1cm}
    \label{fig:cca-vs-opt-number_of_SBS}
    
\end{figure}

We compare CCA with the optimal solution (OPT), where OPT is calculated by solving CCP (as defined by (\ref{eq:obj})-(\ref{eq:constant-int})) exactly. Due to the high complexity of solving OPT (due to the integrality cosntraints), we limit this comparison study to the $20$ highest-popularity videos in the dataset.
We set the local delay $d = 500$ ms, remote delay $D = 2.5$ s and $5$ s. We compare the average delay by varying cache size for each SBS from $5$ GB to $50$ GB with fixed number of SBSs, $N = 6$ (Figure~\ref{fig:cca-vs-opt-cache_size}), and varying the number of SBSs from $6$ to $9$ when the size of each cache is fixed at $25$ GB (Figure~\ref{fig:cca-vs-opt-number_of_SBS}).

We can observe from Figure~\ref{fig:cca-vs-opt-cache_size} that CCA closely approaches OPT at all cache space values. Note that when the cache space is very small, very few videos can be accommodated in the cache, and neither CCA and OPT performs very well. On the other hand, for sufficiently large cache spaces, the entire video set can be store in each caches (or at least in the local network), and both CCA and OPT perform very well. This explains why the the performance of CCA and OPT are very close at these two extremes. 
From Figure~\ref{fig:cca-vs-opt-number_of_SBS}, we see that the ratio of the average delays under CCA and OPT decreases with increase in SBSs. From Figures~\ref{fig:cca-vs-opt-cache_size} and \ref{fig:cca-vs-opt-number_of_SBS}, we see that even in the worst case (within the range of the settings simulated), this performance ratio is within a factor of 2.







\subsection{Effect of Online Implementation Paramters}

Next we implement the online version CCA, with the goal of evaluating the effect of algorithm parameters, $W$ (window size) and $\alpha$ (weighting parameter) on the performance. The performance is measure both in terms of (i) average fetched bytes, (ii) average playout delay. 

The average fetched bytes, which represents the total amount of bytes used by CCA in the local or remote network to serve a video request on an average, consists of two parts: (a) Delivery bytes: The bytes actually fetched from the remote server, or from another cache (SBS) in the local network, at the time of serving a video request; (b) Reoptimation bytes: The bytes transferred over the local network or over the link to the remote server(s) during the re-optimization process under CCA, which happens periodically (end of each time window of length $W$). Accordingly, we plot the average fetched bytes including and excluding reoptimization bytes, where the latter (i.e., when reoptimization bytes are excluded) only consists of the delivery bytes. We also distinguish the fetched bytes based on whether they were fetched over the local network (i.e., from another SBS cache) or from the remote server. 

\begin{figure*}
\begin{minipage}[t]{0.32\textwidth}
  \includegraphics[width=\linewidth]{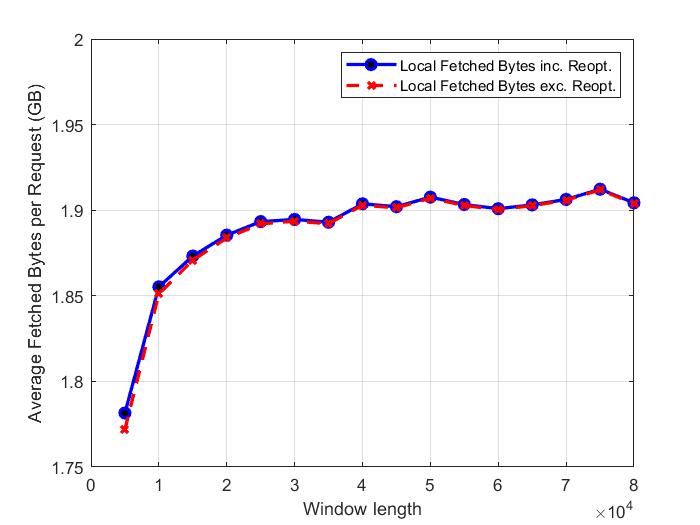}
  \vspace{-0.6cm}
  \caption{Local Fetched Bytes vs Window Length}
  \label{fig:window:first}
\end{minipage}%
\hfill 
\begin{minipage}[t]{0.32\textwidth}
  \includegraphics[width=\linewidth]{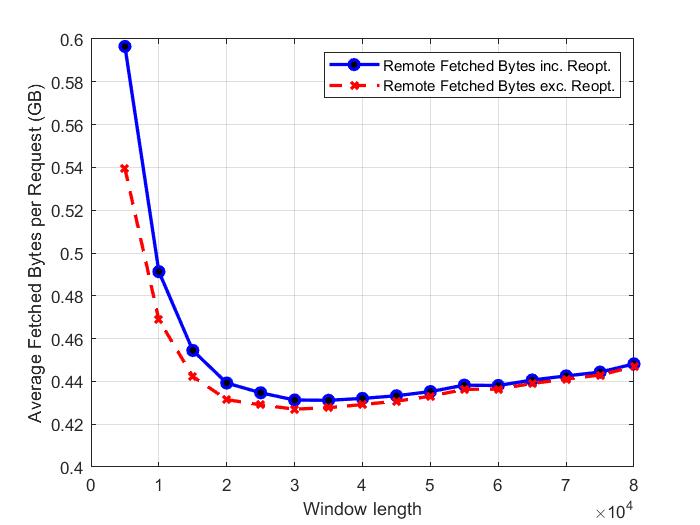}
  \vspace{-0.6cm}
  \caption{Remote Fetched Bytes vs Window Length}
  \label{fig:window:second}
\end{minipage}%
\hfill
\begin{minipage}[t]{0.32\textwidth}
  \includegraphics[width=\linewidth]{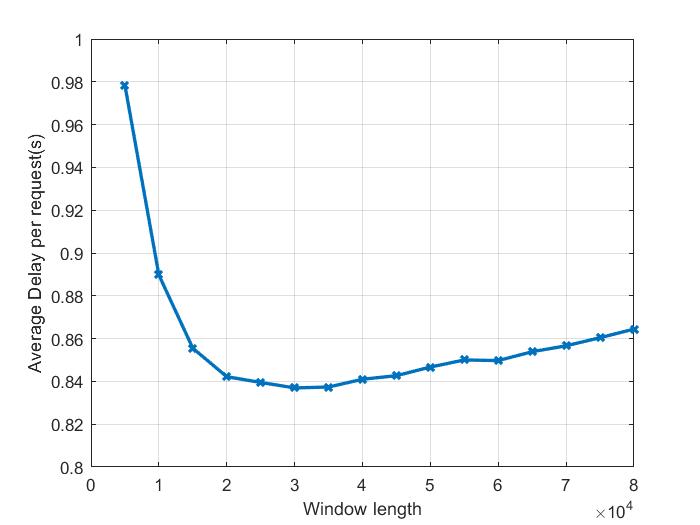}
  \vspace{-0.6cm}
  \caption{Average Delay vs Window Length}
  \label{fig:window:third}
\end{minipage}%
\vspace{-0.4cm}
\end{figure*}

In the following simulations, we set the local delay $d$ = $500$ ms, remote delay = $5$ s, Number of SBSs, $N$ = $4$, Cache space per SBS = $500$ GB, and consider the most popular $3000$ videos from the Netflix Prize Dataset. 
Figures~\ref{fig:window:first}-\ref{fig:window:third} show the variation in fetched bytes (local and remote) and playout delay as the window length $W$ varies, for a fixed $\alpha=0.4$. For these figures, the window length is measured in terms of the total number of requests (across all SBSs), although it could equivalently interpreted in time units as well. We note that the local fetched bytes increases modestly as $W$ increases, before almost flattening out. On the other hand, the remote fetched bytes attain a minimum at an intermediate value of $W$, which is in this case is about $30,000-40,000$. This is intuitively expected: when the window size $W$ is very small, then the estimated $\pi_k$ values do not reflect the true popularity values of the videos. On the other hand, when the window size is very large, any dynamic change in the true popularity values is not immediately reflected in the estimated $\pi_k$ values. From Figure~\ref{fig:window:third}, we note that that the average playout delay for each video is also minimized at about the same window size. Comparing Figures~\ref{fig:window:first}-\ref{fig:window:second}, we see that the local fetched bytes is about 3-5 times the remote fetched bytes. However, since the local network bandwidth is expected to be much larger (at least an order more) than the bandwidth to the remote server(s), our primary network goal is to minimize remote fetched bandwidth (along with minimizing average video playout delay). This is attained at an intermediate value of the window size, which in practice can be determined by analysis of the timescale at which the video popularities change on an average.

\begin{figure*}
\begin{minipage}[t]{0.32\textwidth}
  \includegraphics[width=\linewidth]{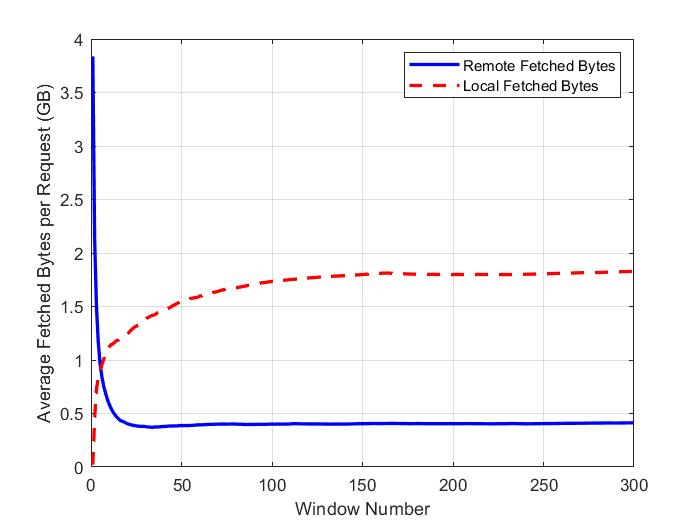}
  \vspace{-0.6cm}
    \caption{Evolution of Total Fetched Bytes (including Reoptimization Bytes) with Window Number.}
    \label{fig:time_evolution}
\end{minipage}%
\hfill 
\begin{minipage}[t]{0.32\textwidth}
  \includegraphics[width=\linewidth]{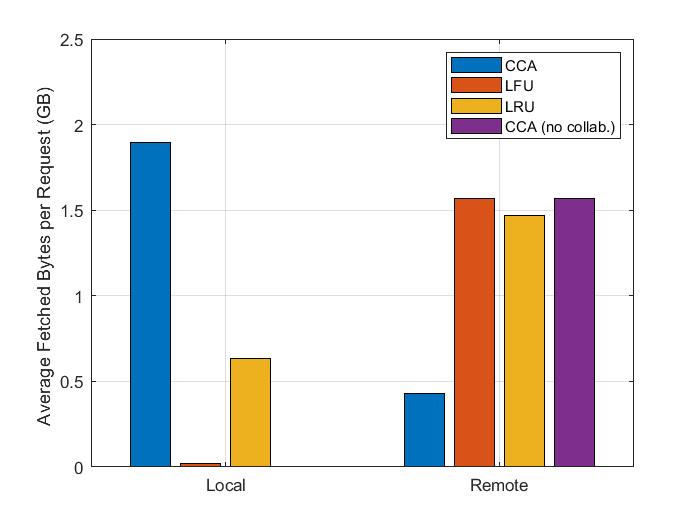}
  \vspace{-0.6cm}
    \caption{Comparison in Total Fetched Bytes with other caching methods.}
    \label{fig:comp-fetched-bytes}
\end{minipage}%
\hfill
\begin{minipage}[t]{0.32\textwidth}
  \includegraphics[width=\linewidth]{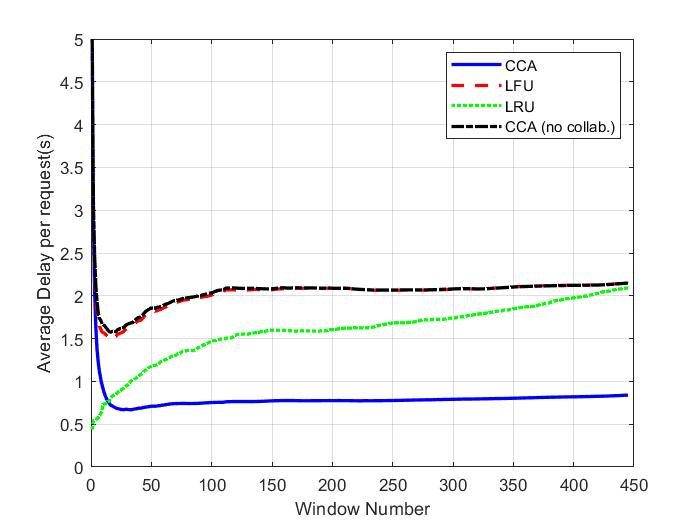}
  \vspace{-0.6cm}
    \caption{Comparison in Average Delay with other caching methods.}
    \label{fig:comp-delay}
\end{minipage}%
\vspace{-0.4cm}
\end{figure*}

The results with variation in $\alpha$ are omitted due to space limitations; however, the performance trends with respect to varying $\alpha$ was observed to be just the opposite of varying $W$, i.e., increasing $\alpha$ produces a trend in the fetched bytes (local and remote) and delay that are similar to those with decreasing $W$. This is expected from intuition: increasing $\alpha$ reduces the weight placed on historical requests, and decreasing $W$ also has a similar effect.  This implies that in practice, we can set one of the parameters to a reasonable value (set $\alpha$ to $0.3-0.5$, say), and choose the other ($W$, say) depending on the timescale of variation in request statistics. 

From Figures~\ref{fig:window:first}-\ref{fig:window:second}, we also note that the amount of reoptimization bytes is fairly small compared to the delivery bytes, for both remote and local delay. In other words, the almost all of the fetched bytes are utilized in directly serving the video requests. This is certainly a desirable feature, and implies that the reoptimization process that that is carried out in the background consumes very little additional bandwidth (both local and remote) but is nevertheless important for adapting the solution to changes in the video popularities.

Figure~\ref{fig:time_evolution} illustrates the temporal variation in average fetched bytes (including reoptimization bytes), for $\alpha = 0.4$ and $W = 35,000$. Here, we assume a cold start, i.e, the caches do not have any cached videos initially. As the window number progresses, videos are cached, and popularity ($\pi_k$) estimates improve, resulting in a sharp drop of remote fetched bytes. Simultaneously, the local fetched bytes increase (finally reaching a steady state value), as most of the requested videos are fetched from a cache in the local network (instead of being fetched from the server), showing the effectiveness of collaborative caching. 






\subsection{Comparison with Other Caching Methods}
\label{sec:comp}

In this section we compare the performance of CCA with two other popularly used caching algorithms, Least Recently Used (LRU) and Least Frequently Used (LFU), extended to include collaborative caching between the SBSs. In the collaborative version of LRU (LFU) that we compare CCA with, each cache implements LRU (LFU) individually, but preferentially obtains a requested video (that is not stored at the cache itself) from one of the other local SBS caches, before requesting the remote server. Also, to quantify the benefits of collaboration, we compare CCA with a version that does not include any collaboration, i.e., Algorithm~\ref{algo:arbitrary-size_caching} is run with Phase I (Greedy Filling) alone.
Figures~\ref{fig:comp-fetched-bytes}-\ref{fig:comp-delay} shows the results. For CCA, we use $W=35,000$ and $\alpha=0.4$, as before. 


From Figure~\ref{fig:comp-fetched-bytes}, we see that compared to LRU and LFU, CCA saves significantly in terms of remote bandwidth. While this saving comes at the cost of increased local network bandwidth use, this is what we desire - since remote bandwidth is expected to be more costly (limited) compared to to local bandwidth. Figure-\ref{fig:comp-delay} shows that CCA provides considerable improvement in video playout delay compared to LRU and LFU, when all the algorithms have a ``cold start''. Furthermore, the average playout delay under CCA reduces quickly after start, and reaches a steady value. For the other two algorithms, the playout delay decreases initially, but then increases slowly but steadily, implying that these algorithms are not able to effectively adapt to changes in the video popularities as time progresses. We also note that CCA with no collaboration performs very close to LFU, as can be expected from intuition.



\section{Conclusion}
\label{sec:Conclusion}
In this paper we considered the problem of collaborative caching of videos across a set of local small-cell base stations (SBSs), which the goal of minimizing video playout delay. Despite the integrality constraints -- and non-convexity of the problem even when the integrality constraints are relaxed -- we provide an algorithm (CCA) that yields a close-to-optimal solution and is computationally efficient. CCA is also implementable in a distributed manner with very limited messaging between the SBSs, and without requiring any explicit coordination between them. Our simulations how that an online implementation of CCA is able to reduce remote bandwidth usage significantly through local sharing of videos among the SBSs. The comparison with collaborative versions of LRU and LFU show that CCA not only results in significantly lower delay, but also settles down (from a cold start) to the lower steady-state playout delay values faster. 

Two extensions of our algorithm may be worth exploring in future work. Firstly, we believe that CCA can be extended to the case where the average local and remote playout delays ($d$ and $D$) are a function of individual video properties (such as the resolution, video type or size), albeit at an increased complexity of Phases II and III of the algorithm. This however needs to be explored further, and formally proven, in future work. Secondly, in our online implementation we have implicitly assumed that the window size parameter $W$ will be chosen/trained manually to match the timescale of variation of the video popularities. Automated and online (re-) training of $W$, which has a significant effect on the performance and adaptiveness of CCA, will be explored in future work. 

\bibliographystyle{IEEEtran}
\bibliography{bib/main}






\appendix

\section*{Proof of Theorem~\ref{th:unit-size}}
We first note that the objective of CCP, as stated in (\ref{eq:obj}), 
\begin{eqnarray}
\sum_k \pi_k  \bigg[ d \sum_i x_{i,k}  \ + \  (D-d)  \sum_i \max_{i'} x_{{i'},k} \bigg], \nonumber
\end{eqnarray}
is a convex maximization (or concave minimization) problem. Note that Phase I (Greedy Filling) maximizes the first term $d \sum_k \pi_k \sum_i x_{i,k}$. It is also easy to observe that each replacement (of video $k_1$ with video $k_2$ inside the while loop) in Phase II (Compare and Replace) improves the objective function in (\ref{eq:obj}). Further note that (\ref{eq:obj}) can be written as 
\begin{eqnarray}
\sum_k \pi_k  \bigg[ d\; n_k  \ + \  N (D-d)\; \min(n_k,1) \bigg], 
\label{eq:obj2}
\end{eqnarray}
which shows that the objective can be expressed just as a function of the variables $n_k$.

We now proceed to prove Theorem~\ref{th:unit-size}. In the following, we assume that all the $\pi_k$ values are distinct. This does not result in any loss of generality: if two values $\pi_k, \pi_{k'}$ are equal, and our (pre-determined) tie-breaking rule prefers $k$ over $k'$, we can increase $\pi_k$ by a sufficiently small amount $\epsilon$ such that $pi_k$ and $\pi_{k'}$ become distinct, and and yet the optimum solution under the new set of $\pi_k$ values remain the same as before. Note that with distinct $\pi_k$ values, the optimum solution to CCP is unique, just as it is under any given tie-breaking rule. 

We divide the set $S_{>1}$ into two subsets: $S^a_{>1}$ and $S^b_{>1}$. The set $S^a_{>1}$ consists of videos in 
$S_{>1}$ for which no copies  have been replaced by Phase II. The set $S^b_{>1}$ consists of videos in 
$S_{>1}$ for which at least one copy have been replaced in Phase II. Thus if $n^I_k$ denote the number of copies of video $k$ populated (across all SBSs) by Phase I, then $S^a_{>1} = \{ k \in S_{>1} | n_k = n^I_k \}$, and $S^b_{>1} = \{ k \in S_{>1} | n_k < n^I_k \}$. Note that since Phase II replaces videos from $S_{>1}$ sequentially from the largest index, $S^b_{>1}$ can can consist of most one video. 

For the sake of contradiction, let us assume that CCA (Algorithm \ref{algo:unit-size_caching}) does not compute an optimum solution, i.e., the objective function value (\ref{eq:obj}) is more in OPT than in the solution computed by CCA. If we let $n^*_k$ be the number of cached copies of video $k$ in OPT, then there exists an index $k'$ such that $n^*_{k'} > n_{k'}$. 
Since $\sum_k n_k = \sum_k n^*_k$ (all of the cache space is utilized by both CCA and OPT), there must be a $k'' \neq k'$ such that $n^*_{k''} < n_{k''}$. 

We consider the four cases separately, depending on whether $k'$ belongs to $S^a_{>1}, S^b_{>1}$, $S_{1}$ or $S_{0}$. 

\noindent \textbf{Case A:} \underline{$k' \in S^a_{>1}$}: There must be a cache $i$ which includes $k'$ under OPT but not under CCA. Since $k' \in S^a_{>1}$, it was not replaced in Phase II of CCA, therefore, all videos $k$ in cache $i'$ must have $\pi_k > \pi_{k'}$, else $k'$ would have been included in cache $i$ in Phase I. Therefore, there must be a video $k'' < k$ which is included in cache $i$ by CCA but not by OPT. We can then improve OPT by replacing video $k'$ in cache $i$ by $k''$, which contradicts the fact that OPT is the optimum solution.

\noindent \textbf{Case B:} \underline{$k' \in S^b_{>1}$}: Again, there must be a cache $i$ which includes $k'$ under OPT but not under CCA. Then there must be a video $k''$ which is included in cache $i$ under CCA, but not under OPT. If $k'' \in S^a_{>1}$, $\pi_{k''} > \pi_{k'}$, and we arrive at a contradiction using an argument similar to Case A. Then $k''$ must be in $S_1$, Since at least one copy of $k'$ must have been replaced in Phase II, the last video in $S_1$, say $k'''$ must satisfy
\begin{equation}
\frac{\pi_{k'''}}{\pi_{k'}} > \frac{d}{ND - (N-1)d}.
\label{eq:case_b}
\end{equation} Since $k'' \geq k'$, $\frac{\pi_{k''}}{\pi_{k'}} > \frac{d}{ND - (N-1)d}$. Then replacing $k'$ in cache $i$ by $k''$ improves OPT, since it changes the objective function in (\ref{eq:obj2}) by $-\pi_{k'} d + \pi_{k''} d + N (D-d) \pi_{k''}$, which is greater than zero, from (\ref{eq:case_b}). This contradicts the fact that OPT is the optimum solution.

\noindent \textbf{Case C:} \underline{$k' \in S_1$}: 
In this case, there must be multiple copies of $k'$ in the local network under OPT, since there is exactly one copy of $k'$ in the local network under CCA. 
Consider the $k''$ for which $n^*_{k''} < n_{k''}$. 
Consider cache $i$ where $k'$ is included under OPT, but not $k''$. If $k'' < k'$, then replacing $k'$ by $k''$ in cache $i$ changes the objective function in (\ref{eq:obj2}) by at least $-\pi_{k'} d +\pi_{k''} d$, which is positive. If $k'' > k'$, then note that $n_{k''} \leq 1$; since $n^*_{k''} < n_{k''}$, $n^*_{k''}$ must be zero. Then replacing video $k'$ by $k''$ in cache $i$ changes the objective function in (\ref{eq:obj2}) in OPT by $-\pi_{k'} d + \pi_{k''} d + N (D-d) \pi_{k''}$. Note that $k''$ is included in $S_1$ under CCA, and therefore must satisfy 
\begin{equation}
\frac{\pi_{k''}}{\pi_{k_1}} > \frac{d}{ND - (N-1)d},
\label{eq:case_c}
\end{equation}
where $k_1$ is the last video in $S_{>1}$. Since $\pi_{k'} < \pi_{k_1}$, it follows that $\frac{\pi_{k''}}{\pi_{k'}} > \frac{d}{ND - (N-1)d}$, which in turn implies that the change $-\pi_{k'} d + \pi_{k''} d + N (D-d) \pi_{k''}$ is greater than zero. This contradicts the fact that OPT is the optimum solution.

\noindent \textbf{Case D:} \underline{$k' \in S_0$}: Again, we consider the $k''$ for which $n^*_{k''} < n_{k''}$. Consider cache $i$ that includes $k'$ but not $k''$. Note that in this case, $k'' \in S_{>1} \cup S_1$. We first consider the case $k'' \in S_1$. In this case, since $n_{k''} =1$, $n^*_{k''} (< n_{k''})$ must be zero. Then replacing $k'$ by $k''$ in cache $i$ changes the objective function (\ref{eq:obj2}) in OPT by at least $\pi_{k''} (d  + N(D-d)) - \pi_{k''} (d  + N(D-d))$, which must be positive, since $k'' < k'$. Now the consider the case $k'' \in S_{>1}$. In this case, replacing $k'$ by $k''$ in cache $i$ changes the objective function (\ref{eq:obj2}) in OPT by at least $-N(D-d) \pi_{k'} - \pi_{k'}d + \pi_{k''}d$. Since $k'$ was not included under CCA, 
\begin{equation}
\frac{\pi_{k'}}{\pi_{k_1}} < \frac{d}{ND - (N-1)d},
\label{eq:case_d}
\end{equation}
for all videos $k_1 \in S_{>1}$. Therefore, we have $\frac{\pi_{k'}}{\pi_{k''}} < \frac{d}{ND - (N-1)d}$, which implies that the change in (\ref{eq:obj2}) due to this replacement, $-N(D-d) \pi_{k'} - \pi_{k'}d + \pi_{k''}d$, is positive. This argues that in both cases of $k'' \in S_1$ and $k'' \in S_{>1}$, the objective function in (\ref{eq:obj2}) under OPT improves by replacing $k'$ in cache $i$ by $k''$, contradicting the fact that OPT is the optimum solution.

We arrive at a contradiction in all of the four cases A, B, C, and D, thus proving that Algorithm~\ref{algo:unit-size_caching} computes the optimum solution to CCP under unit video sizes. 



\section*{Proof of Theorem~\ref{th:arbitrary-size}}
The proof of the result relies on two parts. In the first part, we show that the fractional solution computed at the end of Phase II is an optimal solution to CCP when (\ref{eq:constant-int}) is relaxed, i.e., Phases I and II of Algorithm~\ref{algo:arbitrary-size_caching} solves the fractional version of CCP optimally. The essence of the argument behind this claim is similar to that in the proof of Theorem~\ref{th:unit-size}, and therefore we will only provide a brief outline of this part of the proof.

We then show that in Phase III, the rounding process and the dropping of fractional videos (needed to satisfy integrality constraints (\ref{eq:constant-int})) only adds an $O(\epsilon)$ sub-optimality gap. Note that a naive round down of the solution computed at the end of Phase II would result in a sub-optimality gap that grows as $O(K \epsilon)$, which is undesirable since the number of videos ($K$) can be large. 

Let us define $z_{i,k} = x_{i,k} s_k, \ \forall i, \forall k$. Then the CCP problem in (\ref{eq:obj})-(\ref{eq:constant-int}) can be rewritten as
\begin{eqnarray}
\sum_k \frac{\pi_k}{s_k}  \bigg[ d \sum_i z_{i,k}  \ + \  (D-d)  \sum_i \max_{i'} z_{{i'},k} \bigg].
\label{eq:obj_scaled1}
\end{eqnarray}
\vspace{-0.2in}
\begin{eqnarray}
\mbox{subject to} & \sum_k z_{i,k}  \leq \  C_i, & \forall i, \label{eq:constant-cap_scaled}\\
& z_{i,k} \in \  \{0,s_k\}, & \forall i, \forall k. \label{eq:constant-int_scaled}
\end{eqnarray}
Note that this representation of the general CCP is similar to the CCP problem formation with unit sizes, but with (i) $w_k = \frac{\pi_k}{s_k}$ replacing the popularity values $\pi_k$, (ii) $z_{i,k}$ being constrained to two values, $0$ and $s_k$, instead of $0$ and $1$. This also provides the intuition behind extending Phases I-II of Algorithm~\ref{algo:unit-size_caching} to Phases I-II of Algorithm~\ref{algo:arbitrary-size_caching}.

Since $z_{i,k} \in \  \{0,s_k\}$, $\max_i z_{i,k} = \min (\sum_i z_{i,k}, s_k)$, so (\ref{eq:obj_scaled1}) can be equivalently written as
\begin{eqnarray}
\sum_k w_k  \bigg[ d \sum_i z_{i,k}  \ + \  N (D-d) \min (\sum_i z_{i,k}, s_k) \bigg].
\label{eq:obj_scaled2}
\end{eqnarray}
Now letting $y_k = \sum_i z_{i,k}$, from (\ref{eq:obj_scaled2}) we have that the objective function (\ref{eq:obj_scaled1}) is equivalent to maximizing (compare with (\ref{eq:obj2}))
\begin{eqnarray}
\sum_k w_k \bigg[ d\; y_k  \ + \  N (D-d)\; \min(y_k,s_k) \bigg].
\label{eq:obj2_scaled}
\end{eqnarray}

We now proceed with the first part of the proof, to show that Phases I-II maximize (\ref{eq:obj_scaled2}) (or equivalently maximize (\ref{eq:obj2_scaled})) subject to (\ref{eq:constant-cap_scaled})-(\ref{eq:constant-int_scaled}), but the integrality constraints (\ref{eq:constant-int_scaled}) are relaxed. Therefore, in this relaxed problem, whose optimal objective function value must upper bound the maximum value of (\ref{eq:obj2_scaled})) subject to (\ref{eq:constant-cap_scaled})-(\ref{eq:constant-int_scaled}), the $z_{i,k}$ values can take any continuous value between $0$ and $s_k$; and therefore, $y_k$ can take continuous values between $0$ and $N s_k$. 
The line of reasoning behind this part of the proof is the same as that in the proof of Theorem~\ref{th:unit-size}, but with $y_k$ replacing $n_k$, and the sets $S_{>1}, S_1$ and $S_0$ defined accordingly, as described in Algorithm~\ref{algo:arbitrary-size_caching}. The sets $S^a_{>1}, S^b_{>1}$ are also defined similarly. Thus if $y^I_k$ denote the aggregate size of video $k$ populated (across all SBSs) by Phase I, then $S^a_{>1} = \{ k \in S_{>1} | y_k = y^I_k \}$, and $S^b_{>1} = \{ k \in S_{>1} | y_k < y^I_k \}$.

As in the proof of Theorem~\ref{th:unit-size}, we assume, without loss of generality, that the $\pi_k$ values are all distinct. For the sake of contradiction, let us assume that CCA (Algorithm \ref{algo:arbitrary-size_caching}) does not compute an optimum solution, i.e., the objective function value (\ref{eq:obj_scaled}) is more in OPT than in the solution computed by CCA. If we let $y^*_k$ be the aggregate size of of video $k$ cached in OPT, then there exists an index $k'$ such that $y^*_{k'} > y_{k'}$. Then there must be a cache $i$ such that the amount of video $k$ stored under OPT in the cache ($z^*_{i,k}$) is more than that stored under CCA ($z_{i,k}$). Since under fractional caching, no space is wasted, there must be a $k'' \neq k'$ such that $z^*_{i,k''} < z_{i,k''}$. 
Again, we consider the four cases separately, depending on whether $k'$ belongs to $S^a_{>1}, S^b_{>1}$, $S_{1}$ or $S_{0}$. For each case, the proof replaces an amount $\delta = \min (z^*_{i,k}-z_{i,k}, z_{i,k''}-z^*_{i,k''})$ of video $k'$ in cache $i$ with video $k''$, and shows that it improves the objective function value OPT, thereby arriving at a contradiction. For each case, the argument is similar to the corresponding case (A, B, C and D) in proof of Theorem~\ref{th:unit-size}, and therefore omitted for brevity.

Finally we analyze the rounding process in Phase III. Towards that, we analyze the change in the objective function due to the dropping of the videos during the rounding phase. 
Let $S^i$ be the set of all videos included in cache $i$ (including fractional) in the solution at the end of Phase II, and $z_{i,k}$ be the corresponding video sizes; note that $z_{i,k}$ could be fractional.   
Then the objective function value in (\ref{eq:obj_scaled2}) at the end of Phase II, denoted by $F$, is expressed as
\begin{eqnarray}
F = \sum_i \sum_{k \in S^i} w_k \bigg[ d\; z_{i,k}  \ + \  (D-d)\; \min(\sum_i z_{i,k},s_k) \bigg]. \ 
\label{eq:f_defn}
\end{eqnarray}
Note that $F$ can be written as $F = \sum_i F_i$ where $F_i = \sum_{k \in S^i} w_k \bigg[ d\; z_{i,k}  \ + \  (D-d)\; \min(\sum_i z_{i,k},s_k) \bigg]$.

Let $\hat{S}^i \subseteq S^i$ be the set of videos in cache $i$ after rounding (dropping) step in Phase III. Then the objective function value in (\ref{eq:obj2_scaled}) at the end of Phase III, denoted by $\hat{F}$, is expressed as
\begin{eqnarray}
\hat{F} =  \sum_i \sum_{k \in \hat{S}^i} w_k \bigg[ d\; z_{i,k}  \ + \  (D-d)\; \min(\sum_i z_{i,k},s_k) \bigg]. \
\label{eq:fhat_defn}
\end{eqnarray}
Note that all $z_{i,k}$ values for $k \in \hat{S}^i$ equal $s_k$ (integral videos). If we let $\hat{F}_i = \sum_{k \in \hat{S}^i} w_k \bigg[ d\; z_{i,k}  \ + \  (D-d)\; \min(\sum_i z_{i,k},s_k) \bigg]$, then $\hat{F} = \sum_i \hat{F}_i$. 

Now to understand the difference between the sets $\cup_i S^i$ and $\cup_i \hat{S}^i$, let us list the videos that can be dropped in Phase III. The videos that may get dropped in Phase III can be grouped into four categories:

\noindent \textit{Group A:} \underline{$k' \in S^a_{>1}$}: Note that at the end of Phase I, at most one video in any cache $i$ (the last one to be included in the cache by the greedy filling step) may be fractional. This, if still present at the end of Phase II, could be dropped at the beginning of Phase III. Thus at most one video from the set $S^a_{>1}$ could be dropped from each cache. 

\noindent \textit{Group B:} \underline{$k' \in S^b_{>1}$}: Since Phase II compares and replaces videos one at a time from $S_{>1}$, there is at most one video in this group, over the entire system of caches.

\noindent \textit{Group C:} \underline{$k' \in S_1$}: The space reallocation for videos in $S_1$, carried our Phase III, will drop at most one video per cache (i.e., the video that gets truncated as the cache is being filled).

\noindent \textit{Group D:} \underline{$k' \in S_0$}: Again, since the compare and replace policy considers one video in $S_0$ at a time, there will be at most one fractional video in $S_0$ in the fractional solution at the end of Phase II. Therefore, there is at most one video in this group, over the entire system of caches.

Next we claim that in any cache $i$, at most one video from Group A and Group C categories can get dropped, i.e, two videos - one belonging to Group A and another belonging to Group C - would not be dropped from the same cache $i$. To see this, note that if $k' \in S^a_{>1}$ gets dropped from cache $i$, none of the videos in cache $i$ were replaced in Phase II. Therefore, a video belonging to Group C cannot be dropped from the cache.

Now consider any cache $i$. Let us first assume that a video in Group A gets dropped from the cache in Phase III. From (\ref{eq:f_defn}) and (\ref{eq:fhat_defn}), we see that the difference in the objective function due to this drop, denoted by $\Delta F_i = F_i - \hat{F}_i$, is given by
\begin{eqnarray}
\Delta F_i & = & \ w_{k'} z_{i,k'} + (D-d)\;w_{k'} \min(\sum_i z_{i,k'}, s_{k}) \nonumber\\
& \leq & \ d w_{k'} s_{k'} + (D-d) w_{k'} s_{k'} \nonumber\\
& \leq & \ D w_{k'} s_{\max},
\label{eq:deltaf}
\end{eqnarray}
where $s_{\max} = \max_k s_k$. Note that $\hat{S}^i$ only contains integral videos, and their weights $w_k \geq w_{k'}$ (since the lowest weight video gets dropped). Therefore,
\begin{eqnarray}
    \hat{F}_i & = & \sum_{k \in \hat{S}^i} w_k \bigg[ d\; z_{i,k}  \ + \  (D-d)\; \min(\sum_i z_{i,k},s_k) \bigg]\nonumber\\
    & \geq & w_{k'} \sum_{k \in \hat{S}^i} \bigg[ d\; z_{i,k}  \ + \  (D-d)\; z_{i,k} \bigg]\nonumber\\
    & \geq & D w_{k'}  \sum_{k \in \hat{S}^i} z_{i,k} \ \geq \ D w_{k'} (C_i - s_{\max}). 
    \label{eq:hatf_lb} 
\end{eqnarray}
The last inequality in (\ref{eq:hatf_lb}) follows from the fact that the entire cache is filled up at the end of Phase II, and dropping the video $k'$ only reduces that by $s_{\max}$. 

Next lets assume that the video $k'$  that gets dropped from cache $i$ belongs to Group C. Inequality (\ref{eq:deltaf}) holds in this case as well. Note that the video $k'$ has the lowest weight among all videos in $S_1$ included in the cache. Since the weight of any video in $S_1$ is no greater than the weight of any video in $S_{>1}$, the video $k'$ has the lowest weight among all videos in $S_{>1}$ and $S_1$ included in cache $i$. Therefore, (\ref{eq:hatf_lb}) holds in this case as well. Therefore, considering only the dropped videos in Groups A and C, from (\ref{eq:deltaf}) and (\ref{eq:hatf_lb}), we have 
\begin{eqnarray}
\frac{\Delta F_i}{F_i} \leq \frac{\Delta F_i}{F_i} \leq \frac{s_{\max}}{C_i-s_{\max}}.
\label{eq:deltaf_f_ratio}
\end{eqnarray}
\begin{eqnarray}
\Delta F & = & F - \hat{F} \nonumber\\
& = & \sum_i \Delta F_i \nonumber\\
& \leq & \sum_i \frac{s_{\max}}{C_i-s_{\max}} F_i \nonumber\\
& \leq & \frac{s_{\max}}{C_{\min}-s_{\max}} F,\label{eq:deltaf_total}
\end{eqnarray}
where $C_{\min} = \min_i C_i$. Now let us consider the effect of the dropping the videos in Group B and Group D. Note that there is only one video in each of these groups (across all caches). For the video $k'$ in Group B, since part of the video was replaced by a video $k''$ in $S_1$, we have $w_{k'}/w_{k''} \leq (N(D-d)+d)/d = \rho$ (say). Since $k''$ is the lowest weight video in $S_1$, it follows that $w_{k'}/w_{k'''} \leq \rho$ holds for all videos $k'''$ included in the fractional solution in either $S_{>1}$ or $S_1$. For the video $k'$ in Group D, since the video has the lowest weight among all videos included in the fractional solution, $w_{k'} \leq w_{k'''}$ holds for all videos $k'''$ included in the fractional solution in either $S_{>1}$ or $S_1$.

Therefore, the change in the objective function due to dropping of the two videos in these two groups (B and D), denoted by $\Delta f$, can be bounded as (where $C = \sum_i C_i$):
\begin{eqnarray}
\Delta f \ \leq \ \frac{(\rho+1) s_{\max}}{C-2 s_{\max}} F \ \leq \  \frac{(\rho+1) s_{\max}}{N C_{\min}-2 s_{\max}} F.
\label{eq:deltaf}
\end{eqnarray}
Then, considering the drops of all the videos (a total of $N+2$ videos at most) from Groups A, B, C, D, and letting $\epsilon = s_{\max}/C_{\min}$, from (\ref{eq:deltaf_total}) and (\ref{eq:deltaf}), we have
\begin{eqnarray}
\frac{\Delta f + \Delta F}{F} & \leq & \frac{\epsilon}{1-\epsilon}  + \frac{\frac{\rho+1}{N} \epsilon}{1 - \frac{2\epsilon}{N}}. \label{eq:final_a}
\end{eqnarray}
For $N=1$, Algorithm~\ref{algo:arbitrary-size_caching} concludes by running Phase I, followed by dropping of the last video, if fractional. Therefore, Theorem~\ref{th:arbitrary-size} obviously holds. Therefore, lets consider $N \geq 2$. Since $N\geq 2$, we have
\begin{eqnarray}
\rho + 1 = \frac{N(D-d) + d}{d} + 1 = \frac{ND - (N-2)d}{d} \leq \frac{ND}{d}. 
\label{eq:rho}
\end{eqnarray}
From (\ref{eq:final_a}), using (\ref{eq:rho}) and $N \geq 2$, we have
\begin{eqnarray}
\frac{\Delta f + \Delta F}{F} \ \leq \ \frac{\epsilon}{1-\epsilon}  + \frac{\frac{D}{d} \epsilon}{1 - \epsilon} \ = \ \frac{(\frac{D}{d} + 1) \epsilon}{1-\epsilon}. \label{eq:final_b}
\end{eqnarray}
We make the reasonable assumption that each cache is large enough to hold at least two videos. Therefore, $C_{\min} \geq 2 s_{\max}$, implying $\epsilon \leq 1/2$. Then the upper bound in (\ref{eq:final_b}), $(\frac{D}{d} + 1) \epsilon/(1-\epsilon)$ is upper bounded by $2(\frac{D}{d} + 1) \epsilon$ which is $O(\epsilon)$. The result follows. 

\end{document}